\documentclass[
    ,final            % use final for the camera ready runs
  ]
  {aipproc}

\layoutstyle{6x9}

\begin{document}

\title{VERITAS and H.E.S.S. observations of the
gamma-ray binary HESS J0632+057}

% PACS correspond to X-ray binaries, X-rays and Gamma-rays


\author{P. Bordas (for the H.E.S.S. Collaboration)}{
  address={Institut f\"{u}r Astronomie und Astrophysik, Universit\"{a}t T\"{u}bingen, Sand 1, 72076 T\"{u}bingen, Germany \\}
}
%\url{http://www.mpi-hd.mpg.de/hfm/HESS/}

\author{G. Maier (for the VERITAS Collaboration)}{
  address={DESY, Platanenallee 6, 15738 Zeuthen, Germany \\ }}
%%  address={DESY, Platanenallee 6, 15738 Zeuthen, Germany \\ \url{http://veritas.sao.arizona.edu/}}}

\begin{abstract}

HESS J0632+057 has been recently identified as a new gamma-ray binary system. The source, located in the Monoceros region and
associated with the massive Be star MWC 148, shows variability from radio to very high energy (VHE) gamma-rays, displaying a maximum of
its non-thermal emission about 100 days after periastron passage (at orbital phases $\sim$ 0.3). We present here the results obtained
with the VERITAS and H.E.S.S Cherenkov telescopes spanning a wide time interval from 2004 to 2012. The source is detected at TeV
gamma-rays at a high significance level at phases $\sim$~0.3. We also report for the first time TeV observations belonging to orbital
phases never explored so far. The VHE gamma-ray results are discussed in a multiwavelength context, focusing on contemporaneous
observations obtained with the $\it{Swift}$-XRT.

\end{abstract}

\maketitle

%%%%%%%%%%%%%%%%%%%%%%%%%%%%%%%%%%%%%%%%%%%%
%% MAINMATTER
%%%%%%%%%%%%%%%%%%%%%%%%%%%%%%%%%%%%%%%%%%%%

\section{HESS J0632+057: a new gamma-ray binary}

The VHE gamma-ray source HESS J0632+057 \cite{Aharonian2007} is a new member of the elusive class of gamma-ray binaries
\cite{Hinton2009}. These objects are characterized by a peak in their broad-band energy spectrum at MeV-GeV energies, showing variable
and usually orbitally modulated high-energy emission. All known gamma-ray binaries are high-mass X-ray binaries, consisting of a
compact object orbiting around a massive star of O or Be type. Besides HESS J0632+057, there are only a few binaries clearly identifed
as VHE gamma-ray sources: PSR B1959-63/LS 2883 \cite{Aharonian2005a}, LS 5039 \cite{Aharonian2005b}, LS I +61 303 \cite{Albert2009,
Acciari2011}. In addition, a hint of a VHE flare from Cyg X-1 has been reported \cite{Albert2007}, and VHE emission from a direction
consistent with that of the newly discovered binary 1FGL J1018.6-5856 \cite{Fermi2012} has been recently found \cite{HESS2012}.

%\subsection{Multiwavelength coverage}

%%%%%%%%%%%%%%%%%%%%%%%%%%%%%%%%%%%%%%%%%
% Add here new X-ray analysis
%%%%%%%%%%%%%%%%%%%%%%%%%%%%%%%%%%%%%%%%%

HESS J0632+057 has been repeatedly observed at X-rays: XMM-Newton detected a hard and variable counterpart of HESS J0632+057 at the
position of MWC 148 \cite{Hinton2009}. {\it{Swift}}-XRT observations showed a softer source \cite{Falcone2010} and the presence of
X-ray outbursts each $321\pm 5$ days \cite{Bongiorno2011}. An updated analysis of {\it{Swift}} data including observations up to March
2012 shows a somewhat shorter value of this periodic outburst, each $\sim 315$ days (See Fig. 1). X-ray pulsations from the source were
not found in recent Chandra observations \cite{Rea2011}. At radio wavelengths, VLT and GMRT observations of the source at 5 and 1.28
GHz detected a variable radio counterpart compatible with both the gamma-ray source and the MWC 148 star position \cite{Skilton2009}.
EVN observations further showed spatially extended radio emission during the X-ray outburst, with the emission peak $\approx 10$ times
larger than the orbit size \cite{Moldon2011}. In the optical band, observations with the Liverpool telescope have been used to obtain
radial velocity measures. Fixing the orbital period to 321 days these data suggest a system with an eccentric orbit ($e \approx 0.83$
with periastron occuring at orbital phases $\sim 0.97$ \cite{Casares2012}.

%%%%%%%%%%%%%%%%%%%%%%%%%%%%%%%%%%%%%%%%%%%%
%% Sample figure:
%%
%% The option [height=...] scales the picture to the given height,
%% without it it would be printed at its nominal size
%%%%%%%%%%%%%%%%%%%%%%%%%%%%%%%%%%%%%%%%%%%%

\begin{figure} 
\includegraphics[height=.25\textheight]{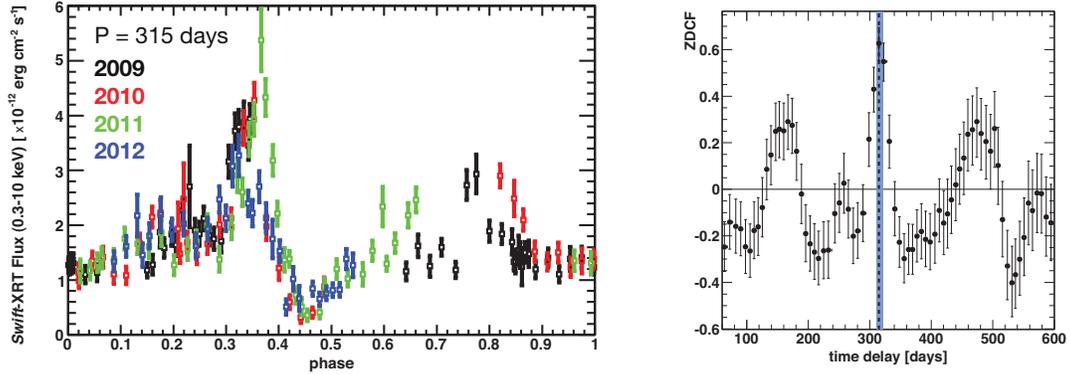} 
\caption{${\it{Left}}$:Phase-folded X-ray lightcurve
using a ${\it{Swift}}$-XRT data-set obtained from 2009 to 2012 (154 compared to 112 flux points in \cite{Bongiorno2011}). The X-ray
peak at phase $\sim0.3$ and the subsequent X-ray dip at phase $\sim0.45$ are clearly seen. ${\it{Right}}$: A z-transformed discrete
cross-correlation function analysis (ZCDF, Alexander 1997) provides an orbital period of P = 315$^{+6}_{-4}$ days} 
\end{figure}

\section{VERITAS and H.E.S.S. observations}

\noindent VERITAS and H.E.S.S. are ground-based gamma-ray observatories, each consisting of four imaging atmospheric Cherenkov
telescopes. The instruments have similar performance, with large effective areas ($> 10^{5}$~m$^{2}$) over a wide energy range (100 GeV to 30 TeV) and
good energy (15-20\%) and angular ($\leq 0.1^{\circ}$) resolutions. The high sensitivity of H.E.S.S. and VERITAS enable the detection of
sources with a flux of 1\% of the Crab Nebula in less than 30 hours of observations.\\

\noindent VERITAS observed HESS J0632+057 for a total of 162 h between 2006 December and 2012 January. About 144 h of data passed
quality selection criteria. All observations were taken with the source at a fixed offset of 0.5$^{\circ}$ from the camera center. The
energy threshold after analysis cuts ranges between 220 GeV and 450 GeV. The most recent observations, taken in 2011/2012, amount to a
total of 34.3 h between November 2011 to January 2012. The analysis of these data provides 163 excess events, implying a detection with
a statistical significance of 9.8 $\sigma$ \\

\noindent H.E.S.S. observed the source yearly from 2004 to 2012. The full dataset consists of 47.3 h of observations, which were
performed over a large range of zenith angles (28$^{\circ}$-58$^{\circ}$). In the 2011/2012 campaign, H.E.S.S. observed the source in
December 2011 and February 2012, providing a total of 8.2 h of good quality data-set. Bad weather prevented however most of the
observations close to the expected peak. We also report here archival H.E.S.S. observations corresponding to orbital phases unexplored
so far. In particular we show new data in the phase range [0.7 - 0.8] (using an orbital period of 315, see above) which were taken in
March 2007, January 2008 and October 2009 and amount to $\sim 7$~h of observing time. Using the {\it{Model analysis}} technique
\cite{Naurois2009} with {\it{standard cuts}}, a $4.1 \sigma$ deviation from the background level is found. A cross-check using a
boosted-decision-tree-based Hillas analysis \cite{Ohm2009} for the same run list, which also makes use of an independent calibration
of the raw data, provides compatible results, with a somewhat higher significance of $4.8 \sigma$ in this phase range.

%%%%%%%%%%%%%%%%%%%%%%%%%%%%%%%%%%%%%%%%%%%%
%% Sample figure:
%%
%% The option [height=...] scales the picture to the given height,
%% without it it would be printed at its nominal size
%%%%%%%%%%%%%%%%%%%%%%%%%%%%%%%%%%%%%%%%%%%%

\begin{figure}
\includegraphics[height=.3\textheight]{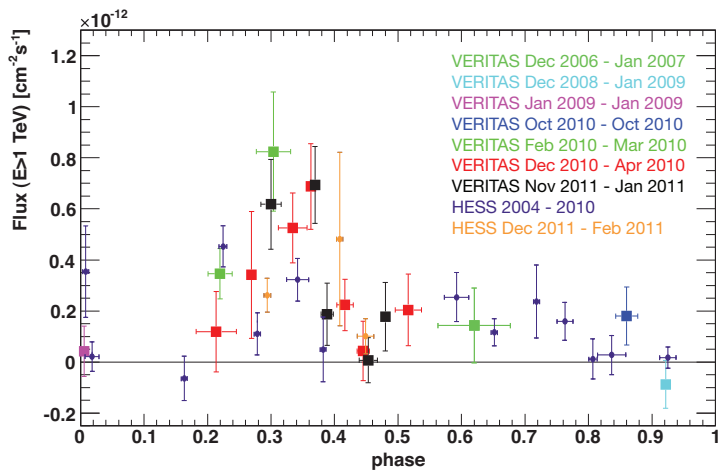}

\caption{VHE fluxes above 1 TeV from H.E.S.S. (purple and orange filled circles) and VERITAS (filled squares;
vertical scale on the left) measurements, folded with the orbital period of 315 days. The marker colors indicate different periods of
observations. Error bars show $1\sigma$ statistical uncertainties.}

\end{figure}

%%%%%%%%%%%%%%%%%%%%%%%%%%%%%%%%%%%%%%%%%%%%
%% SAMPLE TABLE
%%
%% Shows the use of \tablehead and \tablenote
%% macros
%%%%%%%%%%%%%%%%%%%%%%%%%%%%%%%%%%%%%%%%%%%%

\begin{table}
\begin{tabular}{lcccccc}
\hline
  & \tablehead{1}{c}{b}{Obs. time [h]}
  & \tablehead{1}{c}{b}{N$_{on}$}
  & \tablehead{1}{c}{b}{N$_{off}$}   
  & \tablehead{1}{c}{b}{excess}   
  & \tablehead{1}{c}{b}{significance [$\sigma$]}   \\
\hline
VERITAS				&  		& 		& 	    	& 		&  	 \\
All data			& 144.2		& 1525 		& 18310  	& 544		& 15.5 	 \\
2011/2012 			&  34.3		&  367		&  2388		& 163.5		& 9.8 	 \\
\hline
HESS				&  		& 		& 	    	& 		&  	 \\
All data			& 47.3 		& 823 		& 11032  	& 270.0		&  10.4	 \\
2011/2012	        	& 8.2 	 	& 148		& 1787	  	& 53.9		&  4.9   \\
phase $\in [0.7-0.8]$	        & 7.1 	 	& 93		& 1228	  	& 33.9		&  4.1   \\
\hline
\end{tabular}
\caption{VERITAS and H.E.S.S. results for the different data-sets}
\label{tab:a}
\end{table}

\section{Summary and conclusions}

VERITAS and H.E.S.S. observations of HESS J0632+057 provide a wide data-set with more than 190 h of observations covering a large
fraction of the system orbital phases. The source is detected at TeV energies at a high significance level around phase 0.3. Results
are consistent in this phase with observations taken from 2004 to 2012, as well as with recent results reported by the MAGIC
Collaboration \cite{Aleksic2012}. Archival data taken with H.E.S.S show a 4.1$\sigma$ deviation from background events from the
direction of HESS J0632+057 at phases $\sim 0.7-0.8$, or about half an orbital period later than the peak at phase $\sim 0.3$.
Additional TeV data are however required to provide more solid conclusions in this phase range. Finally, an updated X-ray data-set
provides a refined system period of P = 315 $^{+6}_{-4}$ days. In addition, an updated z-CDF analysis (see, e.g. \cite{Alexander1997}) with
the newest data shows a $\sim 4 \sigma$ correlation of the X-ray and VHE fluxes.

%%%%%%%%%%%%%%%%%%%%%%%%%%%%%%%%%%%%%%%%%%%%%%%%
%% BACKMATTER
%%%%%%%%%%%%%%%%%%%%%%%%%%%%%%%%%%%%%%%%%%%%%%%%

\begin{theacknowledgments}
\small{This research is supported by grants from the U.S. Department of Energy Office of Science, the U.S. National Science Foundation and the
Smithsonian Institution, by NSERC in Canada, by Science Foundation Ireland (SFI 10/RFP/AST2748) and by STFC in the U.K. We acknowledge
the excellent work of the technical support staff at the Fred Lawrence Whipple Observatory and at the collaborating institutions in the
construction and operation of the instrument. The support of the Namibian authorities and of the University of Namibia in
facilitating the construction and operation of H.E.S.S. is gratefully acknowledged, as is the support by the German Ministry for
Education and Research (BMBF), the Max Planck Society, the French Ministry for Research, the CNRS IN2P3 and the Astroparticle
Interdisciplinary Programme of the CNRS, the U.K. Science and Technology Facilities Council (STFC), the IPNP of the Charles
University, the Polish Ministry of Science and Higher Education, the South African Department of Science and Technology and
National Research Foundation, and by the University of Namibia. GM acknowledges support through the Young Investigators Program
of the Helmholtz Association. PB acknowledges support from the German Federal Ministry of Economics and Technology through DLR
grant 50 OG 1001. PB also acknowledges the excellent work conditions at the INTEGRAL Science Data Center.}

\end{theacknowledgments}



\bibliographystyle{aipproc}   % if natbib is available

%\bibliography{sample}

%  \typeout{******************************************}
%  \typeout{** Please run "bibtex \jobname" to optain}
%  \typeout{** the bibliography and then re-run LaTeX}
%  \typeout{** twice to fix the references!}
%  \typeout{******************************************}

\end{document}